\documentclass{elsart}

\usepackage{amsmath,amssymb,graphicx,bm}

\newcommand{\be}{\begin{equation}}
\newcommand{\ee}{\end{equation}}
\newcommand{\bea}{\begin{eqnarray}}
\newcommand{\eea}{\end{eqnarray}}
\newcommand{\ba}{\begin{align}}
\newcommand{\ea}{\end{align}}

\newcommand{\lm}{\Lambda}

\newcommand{\Vlk}{V_{{\rm low}\,k}}
\newcommand{\VNN}{V_{\rm NN}}
\newcommand{\Tlk}{T_{{\rm low}\,k}}
\newcommand{\la}{\Lambda}
\newcommand{\kf}{k_{\text{F}}}
\newcommand{\tr}{{\rm Tr}}
\newcommand{\fm}{\, \text{fm}}
\newcommand{\fmi}{\, \text{fm}^{-1}}
\newcommand{\mev}{\, \text{MeV}}
\newcommand{\gevi}{\, \text{GeV}^{-1}}
\newcommand{\vtn}{V_{\text{3N}}}
\newcommand{\veff}{{\overline V_{\text{3N}}}}

\begin{document}

\begin{frontmatter}

\title{Is nuclear matter perturbative with low-momentum interactions?}

\author{S.K. Bogner}$^1$,
\ead{bogner@mps.ohio-state.edu}
\author{A. Schwenk}$^2$,
\ead{schwenk@indiana.edu}
\author{R.J. Furnstahl}$^1$ and
\ead{furnstahl.1@osu.edu}
\author{A. Nogga}$^3$
\ead{a.nogga@fz-juelich.de}
\address{$^1$Department of Physics, The Ohio State University, 
Columbus, OH 43210\\
$^2$Nuclear Theory Center, Indiana University, Bloomington, IN 47408\\
$^3$Institut f\"ur Kernphysik, Forschungszentrum J\"ulich, 
52425 J\"ulich, Germany}

\begin{abstract}

The nonperturbative nature of inter-nucleon interactions is explored
by varying the momentum cutoff of a two-nucleon potential.
Conventional force models, which have large cutoffs, are nonperturbative 
because of strong short-range repulsion, the iterated tensor interaction, 
and the presence of bound or nearly-bound states. But for low-momentum 
interactions with cutoffs around $2\,\mbox{fm}^{-1}$, the softened potential
combined with Pauli blocking leads to corrections in nuclear matter
in the particle-particle channel that are well converged at
second order in the potential, suggesting  that perturbation 
theory can be used in place of Brueckner resummations. 
Calculations of nuclear matter using the low-momentum
two-nucleon force $\Vlk$ with a corresponding leading-order 
three-nucleon (3N) force from chiral effective field theory (EFT)
exhibit nuclear binding in the Hartree-Fock approximation, and
become less cutoff dependent with the inclusion of the dominant 
second-order contributions. The role of the 3N force is essential 
to obtain saturation, and the contribution to the total potential 
energy is compatible with EFT power-counting estimates.     

\end{abstract}

\end{frontmatter}

\section{Introduction}

Conventional wisdom among nuclear physicists, 
as summarized by Bethe in his
review of over 30 years ago~\cite{BETHE71}, holds that successful 
nuclear matter calculations must be
nonperturbative in the inter-nucleon interactions.
The possibility of a soft potential providing a perturbative solution to
the nuclear matter problem was discarded at that time, and saturation firmly
identified with the density dependence due to the tensor force~\cite{BETHE71}.
Subsequent work on the nuclear matter problem \cite{nmrefs}
has not significantly
altered the general perspective or conclusions of Bethe's review
(although the role of three-nucleon (3N) 
forces has been increasingly emphasized).
However, recent results from renormalization-group-based 
low-momentum potentials, coupled with insight from effective 
field theory (EFT), mandate that these conclusions be revisited.
These developments affect not only the technical conclusions but also
the underlying philosophy of nuclear matter calculations. 

Nonperturbative behavior in the
particle-particle channel for nuclear forces arises from several sources.
First is 
a strongly repulsive short-range interaction, 
which requires at least a summation of particle-particle ladder 
diagrams~\cite{BETHE71}.
Second is 
the tensor force, e.g., from pion exchange,
which is highly singular at short distances, and requires iteration
in the triplet channels~\cite{tensor,Fleming}.
Third is the presence of  
low-energy bound states or nearly-bound states, which are found in the
nucleon-nucleon (NN) S-waves. 
These states
imply poles in the scattering $T$ matrix that render the perturbative Born
series divergent.
All of these nonperturbative features are present in conventional
high-precision NN potentials, such as the Argonne $v_{18}$ 
potential~\cite{AV18},
which has been used in the most accurate \textit{ab initio} 
calculations
of nuclei and nuclear matter to date~\cite{GFMC1,GFMC2,AKMAL}.

The philosophy behind 
the standard approach to nuclear matter is to attack these
features head-on.  This attitude  
was succinctly stated by Bethe~\cite{BETHE71}: 
\begin{quote}
``The theory must be such that it can deal with any
NN force, including hard or `soft' core, tensor
forces, and other complications.  It ought not to be necessary to tailor
the NN force for the sake of making the computation of nuclear matter
(or finite nuclei) easier, but the force should be chosen on the basis
of NN experiments (and possibly subsidiary experimental evidence, like
the binding energy of H$^3$).''
\end{quote}
In contrast, the EFT perspective stresses that the potential is not an
observable to be fixed from experiment (there is no ``true potential''), 
but that an infinite number of potentials are capable of
accurately describing low-energy physics~\cite{LEPAGE}.
In order to be predictive and systematic, an organization (``power
counting'') must be
present to permit a finite truncation of possible terms in the potential.
If a complete inter-nucleon potential
is used, including many-nucleon interactions, then all \emph{observable}
results should be
equivalent up to truncation errors.
The EFT philosophy applied to nuclear matter implies
using this freedom to pick a
convenient and 
efficient potential under the conditions of interest.

A particularly useful class of energy-independent NN 
potentials is characterized by a momentum cutoff $\Lambda$, which
limits the resolution of details in the potential. 
One starts with an accurate NN potential (such as
Argonne $v_{18}$~\cite{AV18} or a chiral 
$\mbox{N}^3\mbox{LO}$ potential~\cite{N3LO,N3LOEGM}) 
and a sufficiently high ``bare''
cutoff (all potentials are cut off at high momentum by form factors
or other regulators)
and evolves the cutoff to lower values, while requiring that
observables (e.g., phase shifts)
for external momenta up to the cutoff are unchanged~\cite{Vlowk1,Vlowk2}.
This can be accomplished by renormalization-group (RG) equations that
preserve the half-on-shell $T$ matrix elements (and therefore the
physical on-shell amplitude) or, equivalently, 
by Lee-Suzuki transformations~\cite{Vlowk1,Vlowk2,VlowkRG}
(effective interactions in momentum space have also been derived 
in Ref.~\cite{Epelbaum}).
Varying the cutoff can be used as a powerful tool to study the
underlying physics scales, to evaluate the
completeness of approximate calculations,
and to estimate truncation errors from
omitted higher-order contributions.

These variable-cutoff potentials reveal the resolution or scale 
dependence of the
first two sources of nonperturbative behavior, which are dampened as
high-momen\-tum intermediate states are eliminated.
In free space, the third source of nonperturbative behavior 
remains independent of the cutoff because the pole positions of
shallow bound states that
necessitate fine tuning are physical observables. 
However, this fine tuning is eliminated in the medium at sufficiently
high density.
Our results show that a low-momentum
two-nucleon force (``$\Vlk$'') eliminates all three sources of
nonperturbative physics for bulk properties of nuclear matter.
In short,
a repulsive core is not constrained by phase shifts and is essentially
removed by
even a moderately low-momentum cutoff, the tensor force is tamed
by a sufficiently low cutoff, and the shallow bound
states become perturbative as a result of Pauli blocking.
For cutoffs around $2\,\mbox{fm}^{-1}$, which preserve phase shifts
up to $330 \mev$ laboratory energy,
the Born series in nuclear matter is well converged at second order in
the potential, suggesting that perturbation theory can be used in
place of Brueckner resummations.

While evolving a soft potential from higher momentum is a new
development in nuclear physics~\cite{Vlowk1,Vlowk2},
attempts to use soft potentials
for nuclear matter were common in the 1960's 
and early 1970's~\cite{softpots}.
It had long been observed that a strongly repulsive 
core is not resolved
until eight times nuclear saturation density~\cite{BETHE71}.  Thus,
saturation is \emph{not} driven by a hard core (unlike liquid $^3$He). 
However, these soft potentials were abandoned because they seemed incapable
of quantitatively reproducing nuclear matter properties. 
Their requiem was given by Bethe~\cite{BETHE71}:
\begin{quote}
``Very soft potentials must be excluded because they do not give
saturation; they give too much binding and too high density. In
particular, a substantial tensor force is required.''
\end{quote}
From the EFT perspective, a failure to reproduce nuclear matter
observables should not be interpreted as showing that the low-energy
potential is wrong, but
that it is incomplete (e.g., many-body forces). This 
misconception still persists and has recently led to the conclusion
that low-momentum two-nucleon interactions are ``wrong'' since they 
do not give saturation in nuclear matter and finite nuclei are 
overbound for lower cutoffs~\cite{KUCKEI,FUJII}.  
The missing physics that invalidates this conclusion is many-body
forces, which were completely neglected in those studies.

In a low-energy effective theory, many-body forces are inevitable; the 
relevant question is how large they are.
It is established beyond doubt that for all realistic potentials,
a significant three-body force is 
required to describe light nuclei~\cite{GFMC1,GFMC2,Nogga,NCSM}.
For variable-cutoff potentials,
the three-body (and higher many-body) components of the
potential also evolve with the resolution scale.
A full RG evolution of the combined two- and
three-body potential is not yet available, but can be approximated by
fitting the three-body potential at each cutoff to the form of the
leading-order 3N force from chiral EFT.
The first study of this sort revealed
that the leading three-body components of the 3N force become
perturbative at lower cutoffs~\cite{Vlowk3NF}, which implies 
they are tractable in many-body calculations.
In this paper, we apply these potentials to nuclear matter and
study the convergence properties of low-momentum interactions.

The plan of this paper is as follows.
In Section~2, a variable-cutoff potential is applied to the Born series
for the two-nucleon amplitude in free space and as a function of
density.  The convergence properties as a function of cutoff are
analyzed quantitatively using an approach introduced long ago by
Weinberg.
In Section~3, calculations of nuclear matter using the low-momentum
two-nucleon interaction $\Vlk$ with a corresponding leading-order 3N
force from chiral EFT are presented in the
Hartree-Fock approximation and an approximation to second order.
Nuclear saturation naturally arises at the Hartree-Fock level and becomes less
cutoff dependent with the inclusion of the dominant second-order 
contributions.
Our conclusions are summarized in Section~4 along with a plan for
further calculations of nuclear matter.

\section{Perturbative ladders with low-momentum interactions}

In this section, we demonstrate that two sources
of nonperturbative behavior in the particle-particle channel, 
namely ``hard core''%
\footnote{We use ``hard core'' as a shorthand for
the strong (but not infinite) short-range repulsion in conventional 
nuclear forces.}
scattering and iterated tensor force contributions, 
can be rendered perturbative by using the RG to lower 
the momentum cutoff. We also show that, at sufficiently high density, 
Pauli blocking suppresses the nonperturbative physics due to bound or
nearly-bound states~\cite{Roepke,Friman}. 
In the language of EFT, this means that the
infrared-enhancements that occur in the presence of such states
in free space~\cite{WeinbergIR} are absent in nuclear matter.

In order to keep the presentation self-contained, we briefly review
the RG method used to construct low-momentum
NN interactions (``$\Vlk$'') starting from an arbitrary potential model 
$\VNN$~\cite{Vlowk1,Vlowk2,VlowkRG}. 
In a given partial wave, two-nucleon scattering is described 
by the $T$ matrix, which with standing-wave boundary conditions and
in our conventions ($\hbar = m_{\rm N} = 1$) is given 
by\footnote{This quantity, which is real, is sometimes referred to 
as the $K$ matrix.}
\be
\label{baretmat}
T(k',k;k^2)=\VNN(k',k) + \frac{2}{\pi} \: \mathcal{P} 
\int_{0}^{\infty} q^2 dq \:
\frac{\VNN(k',q) \, T(q,k;k^2)}{k^2-q^2} \,.
\ee
Here $\mathcal{P}$ denotes a principal value integration and the scattering 
phase shifts are given in terms of the diagonal $T$ matrix elements, 
$\tan{\delta(k)} = -k \, T(k,k;k^2)$.
Next, we define a low-momentum version of Eq.~(\ref{baretmat}) in which a
cutoff $\Lambda$ is imposed on loop integrals and the external legs of the 
interaction, 
\be
\label{lowktmat}
\Tlk(k',k;k^2)=\Vlk(k',k) + \frac{2}{\pi} \: \mathcal{P} 
\int_{0}^{\Lambda} q^2 dq \:
\frac{\Vlk(k',q) \, \Tlk(q,k;k^2)}{k^2-q^2} \,.
\ee
As the matching condition, we demand that the half-on-shell $T$ matrices are
equivalent $\Tlk(k',k;k^2) = T(k',k;k^2)$ for $k',k < \lm$ and all
cutoffs. This ensures that
$\Vlk$ gives the same low-momentum two-nucleon observables as $\VNN$
and has the advantage that it avoids energy-dependent interactions,
which are inconvenient in many-body applications.
The cutoff independence of the low-momentum $T$ matrix leads to an RG
equation (for details see~\cite{VlowkRG})
\be
\label{rge}
\frac{d}{d\Lambda} \, \Vlk(k',k) = \frac{2}{\pi} \, \frac{\Vlk(k',\lm) \, 
\Tlk(\lm,k;\lm^2)}{1-(k/\lm)^2} \,.
\ee
Whereas the $T$ matrices in Eqs.~(\ref{baretmat}) and~(\ref{lowktmat}) are
right-side half-on-shell, the $T$ matrix in the RG equation is left-side
half-on-shell and is not RG invariant.
This RG equation is asymmetric in $k$ and $k'$, so it generates 
a non-hermitian $\Vlk$. We avoid 
this by working with a symmetrized version of Eq.~(\ref{rge}), 
\be
\label{symrge}
\frac{d}{d\Lambda} \Vlk(k',k) = \frac{1}{\pi} \biggl( 
\frac{\Vlk(k',\lm) \, 
\Tlk(\lm,k;\lm^2)}{1-(k/\lm)^2} +
\frac{\Tlk(k',\lm;\lm^2) \, \Vlk(\lm,k)}{1-(k'/\lm)^2} \biggr) ,
\ee
which
preserves the on-shell $T$ matrix (i.e., observables) and
is equivalent to the Okubo hermitization of effective interaction theory. 
We note that
$\Vlk$ can be obtained by numerically integrating the RG equation with 
$\VNN$ as the large-cutoff initial condition,
or equivalently by Feshbach projection-operator techniques such as the 
Lee-Suzuki method (for details see~\cite{Vlowk2,VlowkRG}).
    
\subsection{Born series in free space vs.\ in medium}

We can now use the RG to study the convergence properties of free-space 
and in-medium $T$ matrices as we change the resolution scale $\la$. In
perturbation theory, the $T$ matrices are given by the Born series,
\be
T(E) =  V + VG_0(E)V + VG_0(E)VG_0(E)V + \ldots ,
\ee 
where the unperturbed propagator $G_0(E)$ is 
\be
G_0(E) = \frac{1}{E-H_0} \quad \text{(free-space)}\,, \qquad 
G_0(E) = \frac{Q_{\kf}}{E-H_0} \quad \text{(in-medium)} \,,
\ee
and $Q_{\kf}$ is the Pauli-blocking operator that allows scattering
only to unoccupied states above the Fermi momentum $\kf$. In momentum space,
the first and second Born terms are given by
\bea
T^{(1)}(k',k;E) &=& V(k',k) \,, \nonumber \\[2mm]
T^{(2)}(k',k;E) &=& \frac{2}{\pi} \: \mathcal{P} \int q^2 dq \:
\frac{V(k',q) \, V(q,k)}{E-q^2} \quad \text{(free-space)} \,, 
  \label{eq:Bornterms}
  \\[2mm]
T^{(2)}(k',k;E) &=& \frac{2}{\pi} \: \mathcal{P} \int q^2 dq \:
\frac{V(k',q) \, Q(q,P;\kf) \, V(q,k)}{E-q^2-P^2/4} \quad 
\text{(in-medium)} \nonumber \,,
\eea 
where $P$ is the total pair momentum, 
and for S-waves the Pauli-blocking operator 
is given after angular integration by
\be
Q(q,P;\kf) = \langle q, P | Q_{\kf} | q, P \rangle =
\begin{cases}
0 & \text{for $q < \sqrt{\kf^2-P^2/4}$} \\
1 & \text{for $q > \kf + P/2$} \\
\frac{\textstyle q^2+P^2/4-\kf^2}{\textstyle qP} & \text{otherwise} .
\end{cases}
\ee
In the above equations, $V(k',k)$ stands for either 
$\Vlk$ or the input potential 
$\VNN$, and the integration is from 
$0$ to $\Lambda$ or $0$ to $\infty$, respectively.
For simplicity, we neglect self-energy effects and restrict our attention to 
total pair momentum $P=0$ in this section. The dependence on $P$ is generally
very weak, and so our conclusions hold for non-zero $P$ as well.

\begin{figure}[t]
\begin{center}
\includegraphics[height=7.0cm,clip=]{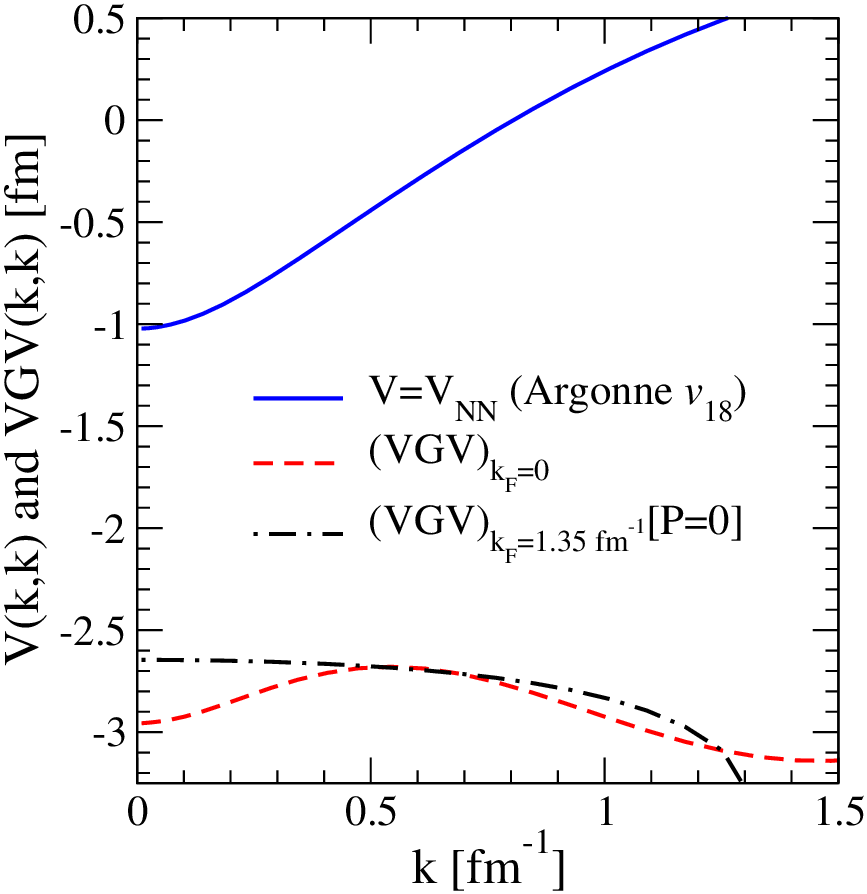}
\hspace*{1mm}
\includegraphics[height=7.0cm,clip=]{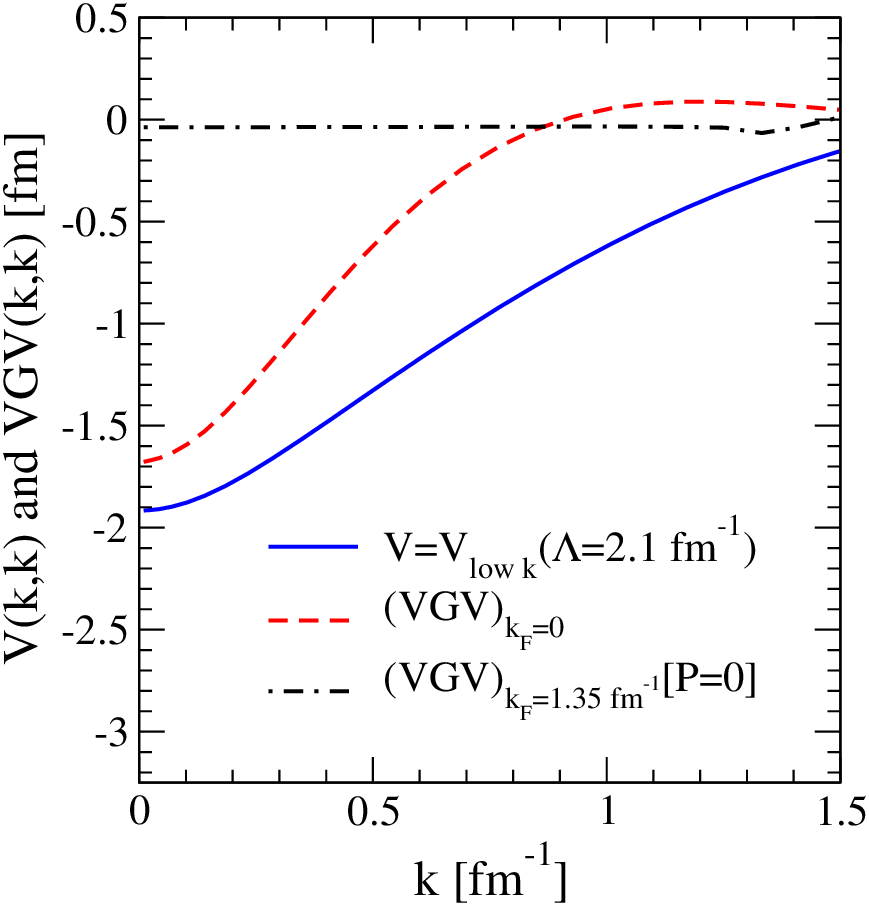}
\end{center}
\caption{\label{fig:V_VGV} 
Perturbative contributions to the diagonal $^1$S$_0$ $T$ matrices 
calculated with the Argonne $v_{18}$ potential (left panel) and with
$\Vlk$ for $\Lambda = 2.1 \fmi$ (right panel). The first-order 
interactions are given by the solid lines. The second-order free-space 
(in-medium) contributions are denoted by the dashed (dot-dashed) lines.
The in-medium contributions are evaluated at saturation density and for
total pair momentum $P=0$. Here and in the following $\Vlk$ is calculated
from the Argonne $v_{18}$ potential.}
\end{figure}

The left panel of Fig.~\ref{fig:V_VGV} shows the first- and second-order 
contributions to the diagonal $T$ matrix elements calculated from the 
Argonne $v_{18}$ potential in the $^1$S$_0$ channel. It is evident that
perturbative expansions of the free-space and in-medium $T$ matrices are 
not possible, as the second-order contribution is several times larger 
than the first-order contribution over all momenta considered. 
This divergence continues in higher orders.
The strong repulsive core in the Argonne $v_{18}$ interaction, or any
other conventional NN potential model, leads to strong high-momentum 
matrix elements, and the momentum-independent nonperturbative behavior
observed in Fig.~\ref{fig:V_VGV} is a signature of ``hard core'' scattering.
We emphasize that cores are not constrained by fits to low-energy two-nucleon
data, and therefore ``hard core'' divergences of perturbation theory 
are not dictated by physics.

Pauli blocking does not change this picture, and in nuclear matter
the second-order contribution is comparable to the free-space one,
as can be seen from Fig.~\ref{fig:V_VGV}. Therefore, ``hard core'' 
scattering always dominates.
One might have expected that the second-order contribution is small
at threshold in nuclear matter, since Pauli blocking suppresses 
the infrared enhancements that occur in free space at low energies. 
Clearly, this is not the case with conventional potential models 
due to the cores.

Referring to the right panel of Fig.~\ref{fig:V_VGV}, we find that the
results obtained from $\Vlk$ with $\Lambda = 2.1 \fmi$ exhibit a very
different pattern. We first observe that the second-order contributions
to the amplitudes are smaller than the first-order Born term over the 
range of momenta considered. The 
second-order free-space contribution is largest at zero energy due to
the large $^1$S$_0$ scattering length,\footnote{With the normalization 
used here, the contributions at $k=0$ will sum to the $^1$S$_0$ neutron-proton 
scattering length of $a_{^1\text{S}_0} = -23.7 \fm$.} while it decreases
rapidly away from threshold, indicating that ``hard core'' scattering 
is absent. We emphasize that the diagonal free-space $T$ matrix elements 
obtained from an exact solution using either the Argonne $v_{18}$ 
potential or $\Vlk$ are identical up to $330 \mev$ lab energy. 
This demonstrates that the 
convergence behavior of the Born series in free space is strongly 
scale-dependent, with lower cutoffs being advantageous.

For the in-medium calculation, we find the striking result 
that the second-order contribution 
is very small over all momenta. This dramatic 
improvement of in-medium convergence can be easily understood from a
simple phase-space argument. In the present calculation for total pair 
momentum $P=0$, the limits of the intermediate-state integrals 
in Eq.~(\ref{eq:Bornterms}) are 
from $\kf$ to $\lm$ for $\Vlk$. In contrast, the loop integrals for 
conventional potentials require large upper limits of $15\mbox{--}20 \fmi$.
Therefore, the restriction of phase-space 
substantially reduces the size of higher-order corrections. 
This reduction is also aided by weaker interactions at higher 
momenta in the dominant S-waves.
   
\begin{figure}[t]
\begin{center}
\includegraphics[width=7.0cm,clip=]{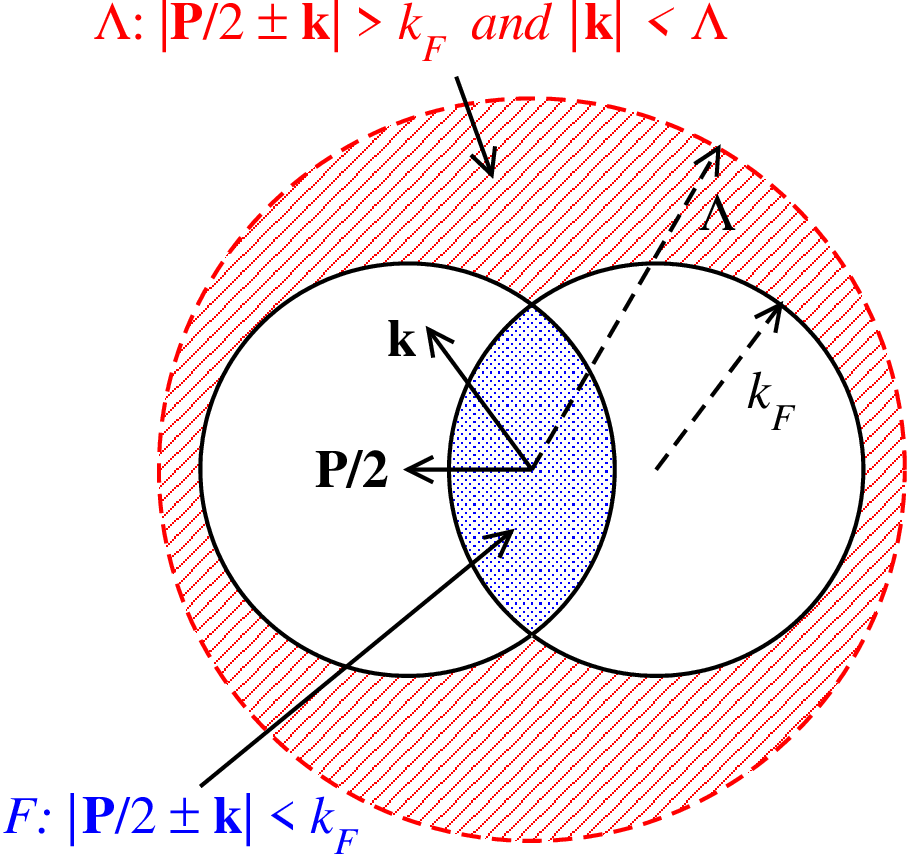}
\end{center}
\caption{\label{fig:fermisphere} 
Pauli-allowed intermediate-state phase space (hatched region 
``$\Lambda$'') for scattering in the particle-particle channel. 
Nucleons are initially inside the Fermi sea (shaded region ``F'') 
and interactions are restricted to momenta below the cutoff $\Lambda$.}
\end{figure}

This phase-space argument also holds for general $P\ne0$.
In Fig.~\ref{fig:fermisphere}, we show the allowed phase space 
for the general case, where a pair of particles with total 
momentum $P$ initially in the Fermi sea (the shaded region ``$F$'') 
scatters to Pauli-allowed intermediate state (the hatched region 
``$\Lambda$''). It is clear from Fig.~\ref{fig:fermisphere} that 
lowering the cutoff suppresses the intermediate-state phase space 
for $P\ne0$ case as well. 
When $P$ is very close to $2\kf$ the suppression is less extreme, but
a corresponding $\Vlk \, G \, \Vlk$ curve as in Fig.~\ref{fig:V_VGV} 
is still less than
$0.3\,\mbox{fm}$ in magnitude throughout the range of $k$.

However, the cutoff also has to be greater than the largest
relative momentum with non-negligible contributions to the
observable of interest. It is evident from Fig.~\ref{fig:fermisphere} that 
for large total pair momentum (e.g., $P=2\kf$), cutoffs 
$\lm < 2 \kf$ cut into the Fermi spheres. For general interactions,
taking the cutoff too small could 
complicate the task of maintaining cutoff-independent observables.
This is because the reshuffling of contributions from
loop integrals to the effective interaction in the RG equation
is defined in free space. In this paper we take a pragmatic 
approach to this issue. Energy-per-particle results for $\lm < 2 \kf$
are found to be very weakly dependent on the cutoff, which
indicates that such configurations contribute negligibly to bulk
properties.%
\footnote{We also note that observables for which
relative momenta $k > 2 \fmi$ contribute significantly will
be model dependent. This is the case, for example, for P-wave pairing
gaps for neutron Fermi momenta $\kf > 2 \fmi$
\cite{Pgaps}.}
Moreover, region ``F'' vanishes as $P \to 2\kf$ and the average 
pair momentum in the Fermi sea is $P_{\text{av}} = \sqrt{6/5} \, \kf$.
Therefore, it is 
likely that the limit $\lm \geqslant 2 \kf$ set by the $P = 2\kf$ 
configurations is too conservative.

In summary, we find
that the perturbative convergence of two-nucleon scattering
is improved dramatically when we use the
RG to evolve the interaction down to lower cutoffs. The resulting
$\Vlk$ is by construction phase-shift equivalent to the
input potential, but $\Vlk$ removes the difficulties due
to repulsive cores. In-medium ladders
become perturbative because lower cutoffs restrict the 
intermediate-state phase-space integrations, and because
NN scattering in the dominant S-waves is weaker
at higher momenta. 

\subsection{Weinberg eigenvalue analysis}

The preceding conclusions can be made rigorous by adopting a method 
introduced by Weinberg in Ref.~\cite{Weinberg} that provides
quantitative conditions for the perturbative convergence 
of the $T$ matrix.  This allows us to identify the cutoffs 
at which ``hard core'' scattering and iterated tensor 
contributions become perturbative. Moreover, the analysis
shows that Pauli-blocking removes nonperturbative behavior
due to low-energy bound or nearly-bound states. Thus, we will
establish as a main result of this paper that the
in-medium $T$ matrix is perturbative over a wide range 
of cutoffs and densities.

Following Weinberg~\cite{Weinberg}, we study the spectrum of
the operator $G_0(z) V$ for complex energies $z$
\be
\label{Weinberg1}
G_0(z) V \, | \Psi_{\nu}(z) \rangle = \eta_{\nu}(z) \, 
| \Psi_{\nu}(z) \rangle \,,
\ee
where the index $\nu$ labels the discrete eigenvalues and eigenvectors.
As demonstrated in~\cite{Weinberg}, the perturbative Born 
series for $T(z)$ diverges if and only if there is an
eigenvalue with $|\eta_{\nu}(z)| \geqslant 1$. The necessity of 
this condition is clear, since 
\be
\sum_{n=0}^{\infty} V \bigl( G_0(z) V \bigr)^n \, | \Psi_{\nu}(z) \rangle =
\sum_{n=0}^{\infty} \bigl( \eta_{\nu}(z) \bigr)^n \, V \, 
| \Psi_{\nu}(z) \rangle \,,
\ee
and thus
\be
T(z) \, | \Psi_{\nu}(z) \rangle = \bigl( 1 + \eta_{\nu}(z)
+ \eta_{\nu}(z)^2 + \ldots \bigr) \, V \, 
| \Psi_{\nu}(z) \rangle
\ee
diverges for $|\eta_{\nu}(z)| \geqslant 1$. Furthermore, 
Weinberg also demonstrates that the rate of perturbative 
convergence is controlled by the largest $|\eta_{\nu}(z)|$, 
with smaller values implying faster convergence.
  
A rearrangement of Eq.~(\ref{Weinberg1}) gives a simple interpretation 
of the eigenvalues $\eta_{\nu}(z)$ in terms of the Schr\"odinger equation,
\be
\bigl( H_0 + \frac{1}{\eta_{\nu}(z)} \, V \bigr) \, | \Psi_{\nu}(z) 
\rangle = z \, | \Psi_{\nu}(z) \rangle .
\ee
The eigenvalue $\eta_{\nu}(z)$ can thus be viewed as an energy-dependent 
coupling constant that must divide $V$ to produce a solution to the 
Schr\"odinger equation at energy $z$. 
If $V$ supports a bound state at 
$z=B < 0$, then there is some $\nu$ with $\eta_{\nu}(B)=1$, which implies 
a divergence of the Born series for nearby energies. However, what 
matters for convergence at a given energy $z$ is not simply the presence 
of nearby physical bound states, but rather the complete set of eigenstates 
that can be shifted to $z$ when the interaction is divided by 
$\eta_{\nu}(z)$.  At negative energies, a purely attractive 
$V$ gives positive $\eta_{\nu}(z)$ values, while a purely repulsive $V$ 
gives negative values, as the sign of the interaction must be flipped
to support a bound state. For this reason, we follow convention and 
refer to negative eigenvalues as repulsive and positive ones as 
attractive. In the case of conventional NN interactions, the repulsive 
core generates at least one large and negative eigenvalue that causes 
the Born series to diverge. 

\begin{figure}[t]
\begin{center}
\includegraphics[width=10.0cm,clip=]{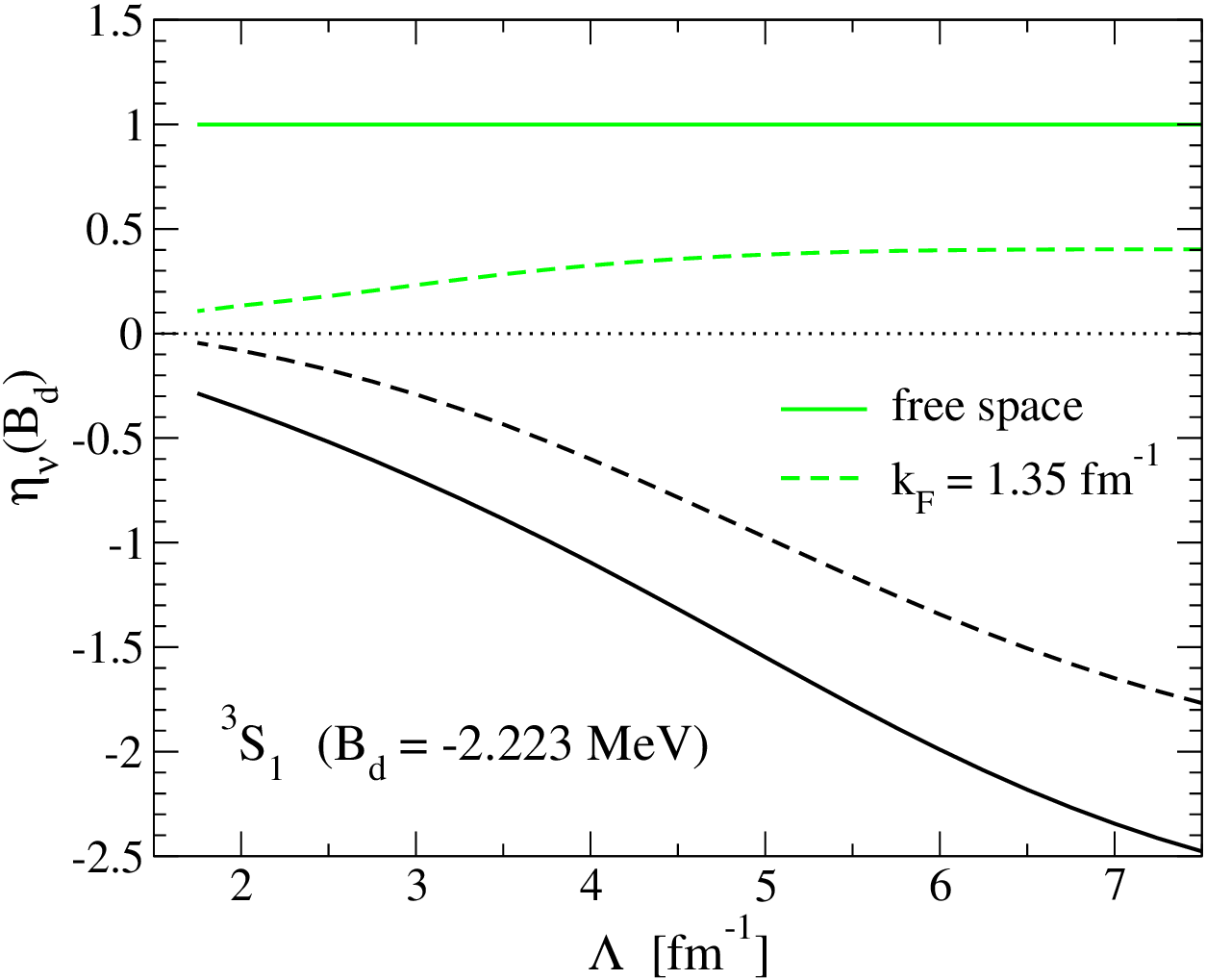}
\end{center}
\caption{\label{fig:weinberglambda} 
Evolution of the two largest Weinberg eigenvalues with $\Lambda$ in 
the coupled $^3\mbox{S}_1$--$^3\mbox{D}_1$ 
channel in free-space (solid) and at saturation
density (dashed). The lighter curves correspond to attractive
eigenvalues and the darker curves to repulsive eigenvalues.}
\end{figure}

\begin{figure}[t]
\begin{center}
\includegraphics[width=10.0cm,clip=]{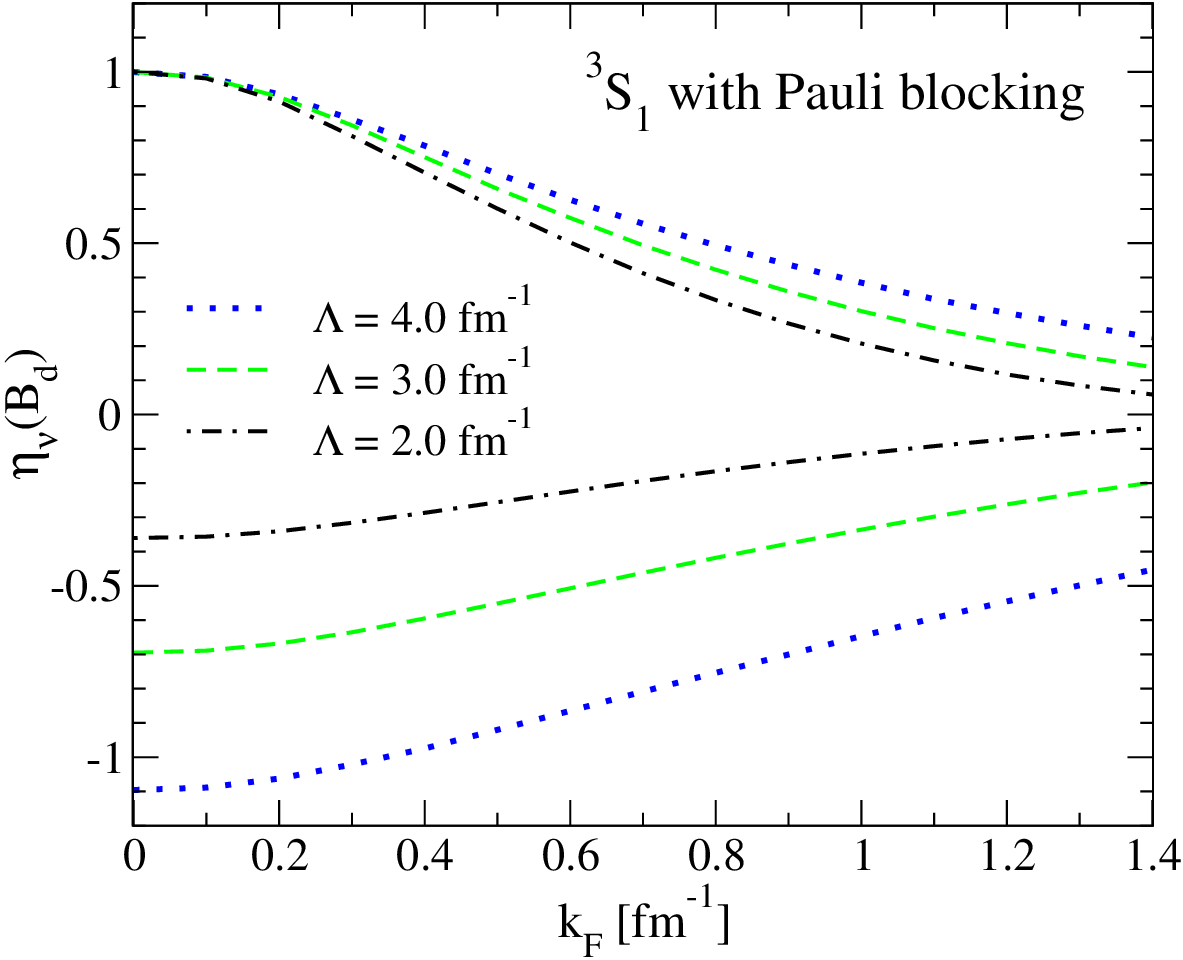}
\end{center}
\caption{\label{fig:weinbergmedium} 
Dependence  on density
of the two largest Weinberg eigenvalues in the 
coupled $^3\mbox{S}_1$--$^3\mbox{D}_1$ 
channel.}
\end{figure}

We apply the spectral analysis to the coupled $^3\mbox{S}_1$--$^3\mbox{D}_1$ 
channel and show 
in Fig.~\ref{fig:weinberglambda} the evolution of the two largest 
Weinberg eigenvalues evaluated at the deuteron pole $z=B_d$, as we lower
the cutoff. 
From the free-space curves (solid lines), we observe 
that the repulsive eigenvalue associated with the 
core and the tensor force is initially large 
but decreases with cutoff, with $|\eta_\nu(B_d)| < 1$ 
for $\Lambda < 3.5 \fmi$. In contrast, the physical divergence 
corresponding to the deuteron pole remains invariant, $\eta_\nu(B_d)=1$, 
as it should. This suggests that the nonperturbative behavior due to
the strong core and strong tensor force is only present for large-cutoff
interactions.

The finite-density curves (dashed lines) show a similar dependence
on cutoff, but now the nonperturbative attractive eigenvalue 
associated with the deuteron in free space is tamed by 
Pauli-blocking effects. This result is general, as illustrated 
by Fig.~\ref{fig:weinbergmedium}, where we show the density 
dependence of the Weinberg eigenvalues for various cutoffs. 
As the density increases, Pauli-blocking drives the largest
attractive Weinberg eigenvalue to a perturbative regime.

It is important to note that the repulsive eigenvalue is 
significantly reduced when the cutoff is lowered from $\Lambda 
= 3 \fmi$ to $\lm = 2 \fmi$. 
Second-order tensor contributions strongly excite 
intermediate-state momenta peaked about $k \approx 2.5\mbox{--}3.0 
\fmi$ in nuclear matter~\cite{GerryMBbook}. Consequently, 
when the cutoff is lowered below this range, the strength
of the tensor force decreases, leading to a smaller Weinberg 
eigenvalue. Chiral EFT interactions have
Weinberg eigenvalues similar to the $\lm =3 \fmi$ curves,
and thus their in-medium convergence properties can be 
improved by using the RG to run their cutoff to lower
values.

We therefore reach a promising conclusion: Large-cutoff sources of 
nonpertur\-bative behavior can be eliminated using 
the RG to lower the momentum cutoff, while the physical source of 
nonperturbative behavior due to bound and nearly-bound states
is suppressed in the medium by the Pauli principle. For simplicity,
the above analysis was for NN interactions only, but 
as we show in the next section, a 
consistent treatment of many-nucleon forces does not change the
general conclusion. 

\section{Nuclear matter results}

It is evident from the preceding section that for nuclear interactions 
with large
cutoffs, the strong cores and tensor force necessitate a resummation 
of particle-particle scattering in nuclear matter. This resummation is 
typically carried out in the Brueckner approach, which sums iterated 
particle-particle scattering within a self-consistent 
mean-field (for reviews see~\cite{Morten,Baldorev}).

\begin{figure}[t]
\begin{center}
\includegraphics[scale=0.5,clip=]{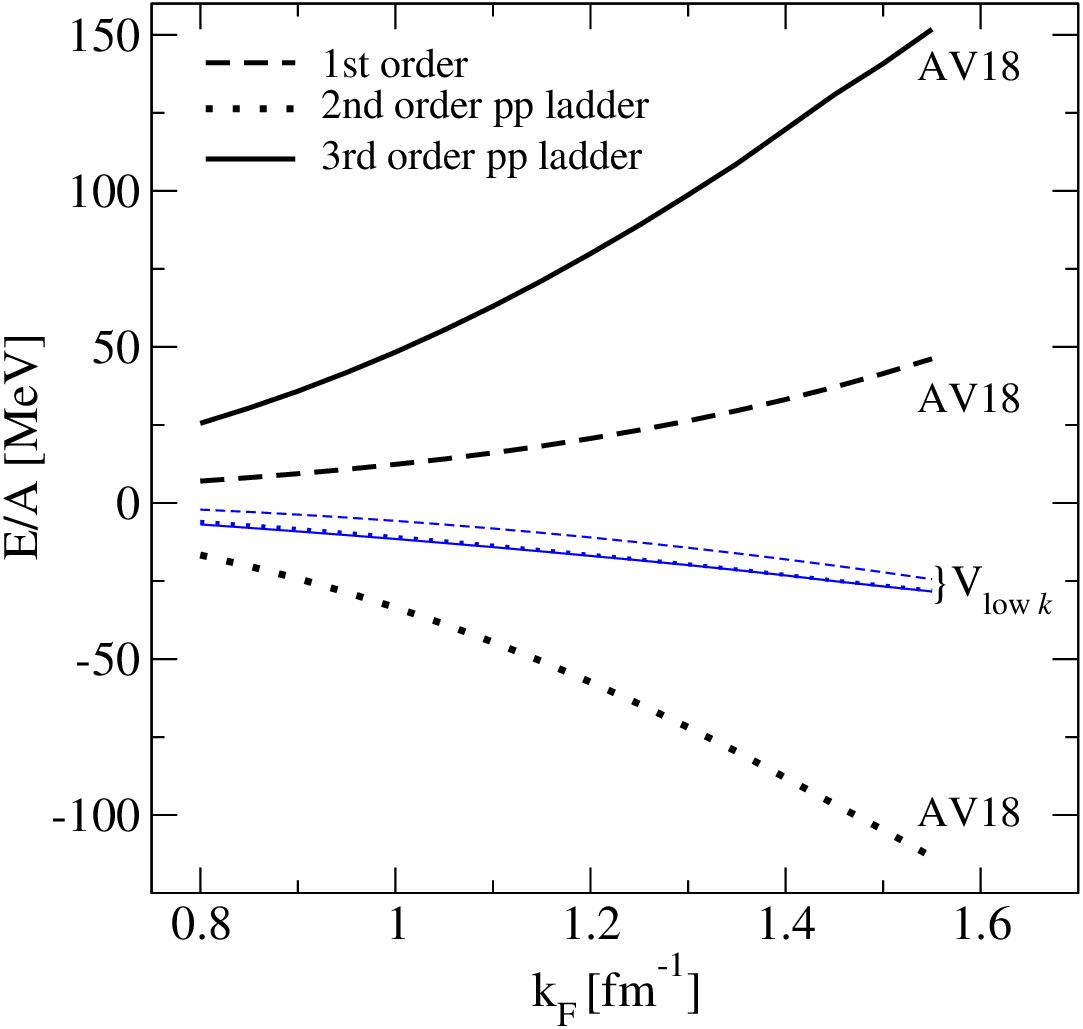}
\end{center}
\caption{Particle-particle contributions to the energy per particle
in symmetric nuclear matter for the Argonne $v_{18}$ potential (thick
lines) and the low-momentum interaction $\Vlk$ with $\lm=2.1 \fmi$ 
(thin lines).}
\label{pert2nf}
\end{figure}

In Fig.~\ref{pert2nf}, we show the particle-particle contributions
to the energy of symmetric matter in perturbation theory. The calculations
are carried out using a continuous single-particle spectrum with 
the effective mass approximation, and all intermediate-state phase-space 
integrations are angle-averaged.
Fig.~\ref{pert2nf} demonstrates that for large-cutoff
interactions, such as the Argonne $v_{18}$ potential, iterated 
particle-particle scattering is nonperturbative. We repeat that 
this behavior is not required by nucleon-nucleon scattering data, and 
may be viewed as an artifact of the large cutoff in the Argonne $v_{18}$ 
potential. In stark contrast, for the low-momentum interaction 
$\Vlk$ with $\lm=2.1 \fmi$ we observe that particle-particle 
ladders are perturbative in nuclear matter. These results are due to
restricted phase space and follow from the behavior of the Weinberg 
eigenvalues, as discussed in the previous section.

With two-nucleon low-momentum interactions only, particle-particle 
correlations do not lead to saturation within the density range where 
nuclear forces are well-constrained experimentally. We also note that 
for two-nucleon forces with cores, saturation occurs at Fermi momenta 
$\kf > 1.5 \fmi$ in the Brueckner 
approach~\cite{Morten,Baldorev,Lejeunenm}, in contrast to 
the empirical value of $\kf = 1.35 \fmi$. Such calculations
overbind nuclear matter by several MeV relative to the empirical value of 
$E/A = -16 \mev$.
Moreover, it is well-established that 3N interactions present important 
contributions in light nuclei~\cite{GFMC2,Nogga,NCSM}. 
In Ref.~\cite{Vlowk3NF}, we showed that 3N contributions
are smaller for low-momentum interactions, and that 3N forces are
required to guarantee regulator-independent results.
Therefore, it is imperative that 3N forces are included in 
calculations of nuclear matter. 

As discussed in the introduction, an RG evolution of the combined 
two- and three-nucleon interactions is not yet available. However, 
we have argued that this evolution can be approximated by fitting 
the leading-order 3N force from chiral EFT at each cutoff to be 
used with $\Vlk$~\cite{Vlowk3NF}. 
This is motivated by the fact that $\Vlk$ becomes independent of
the starting two-nucleon potential as the cutoff 
is lowered to $\la \lesssim 2 \fmi$.
The same model-independent $\Vlk$ is obtained by evolving
a chiral $\mbox{N}^3\mbox{LO}$ two-nucleon interaction to
smaller cutoffs. The RG evolution induces contact interactions which 
are of higher order in a chiral expansion and are necessary to maintain
cutoff-independent two-nucleon observables.  
This indicates that $\Vlk$ effectively parameterizes a 
chiral EFT interaction with higher-order contact terms, and supports 
our assumption that the low-momentum 3N interaction corresponding to $\Vlk$ is 
well-approximated by the leading-order chiral 3N force.  

Therefore, we have augmented $\Vlk$ with the leading-order 
chiral 3N force~\cite{Vlowk3NF}. The latter contains a long-range
 $2 \pi$-exchange part $V_c$, an intermediate-range $1 \pi$-exchange 
part $V_D$ and a short-range contact interaction 
$V_E$~\cite{chiral3NF1,chiral3NF2}. The $2 \pi$-exchange interaction is 
given by
\be
V_c = \frac{1}{2} \, \biggl( \frac{g_A}{2 f_\pi} \biggr)^2 
\sum\limits_{i \neq j \neq k} 
\frac{({\bm \sigma}_i \cdot {\bf q}_i) ({\bm \sigma}_j \cdot 
{\bf q}_j)}{(q_i^2 + m_\pi^2) (q_j^2 + m_\pi^2)} \: F_{ijk}^{\alpha\beta} \,
\tau_i^\alpha \, \tau_j^\beta \,,
  \label{eq:Vc}
\ee
where ${\bf q}_i = {\bf k}'_i - {\bf k}_i$ denotes the difference of initial
and final nucleon momenta and
\be
F_{ijk}^{\alpha\beta} = \delta^{\alpha\beta} \biggl[ - \frac{4 c_1 
m_\pi^2}{f_\pi^2} + \frac{2 c_3}{f_\pi^2} \: {\bf q}_i \cdot {\bf q}_j
\biggr] + \sum_\gamma \, \frac{c_4}{f_\pi^2} \: \epsilon^{\alpha\beta\gamma}
\: \tau_k^\gamma \: {\bm \sigma}_k \cdot ( {\bf q}_i \times {\bf q}_j )
   \,,
\ee
while the $1 \pi$-exchange and contact interactions are given respectively by
\begin{align}
V_D &= - \frac{g_A}{8 f_\pi^2} \, \frac{c_D}{f_\pi^2 \la_\chi}
\: \sum\limits_{i \neq j \neq k} 
\frac{{\bm \sigma}_j \cdot {\bf q}_j}{q_j^2 + m_\pi^2} \: ({\bm \tau}_i
\cdot {\bm \tau}_j) \, ({\bm \sigma}_i \cdot {\bf q}_j) \,, \\
V_E &= \frac{c_E}{2 f_\pi^4 \la_\chi} \: \sum\limits_{i \neq
j \neq k} ({\bm \tau}_j
\cdot {\bm \tau}_k) \,.
  \label{eq:VDVE}
\end{align}
In applying Eqs.~(\ref{eq:Vc})--(\ref{eq:VDVE}), 
we use $g_A = 1.29$, $f_\pi = 92.4 \mev$ and $m_\pi = 138.04 \mev$
and the $c_i$ constants extracted by the Nijmegen group in a partial
wave analysis with chiral $2 \pi$-exchange~\cite{const}. These are
$c_1 = -0.76 \gevi$, $c_3 = -4.78 \gevi$ and $c_4 = 3.96 \gevi$. 
The fit values for the $c_D$ and $c_E$ low-energy constants
are tabulated in Ref.~\cite{Vlowk3NF} for $\la_\chi = 700 \mev$.

Motivated by the above discussion, we proceed to calculate the energy
per particle in symmetric nuclear matter from $\Vlk$ and $\vtn$ 
in many-body
perturbation theory. The $\Vlk$ and $\vtn$ Hartree-Fock contributions are
given by
\begin{align}
\frac{E^{(1)}_{\Vlk}}{V} &= \frac{1}{2} \: \tr_{\sigma_1, \tau_1} 
\tr_{\sigma_2, \tau_2} \int \frac{d{\bf k}_1}{(2\pi)^3}
\int \frac{d{\bf k}_2}{(2\pi)^3} \: n_{k_1} \, n_{k_2} \,
\langle 1 2 \, | \, \Vlk \, (1-P_{12}) \, | \, 1 2 \rangle \,,\\
\frac{E^{(1)}_{\vtn}}{V} &= \frac{1}{6} \: \tr_{\sigma_1, \tau_1} 
\tr_{\sigma_2, \tau_2} \tr_{\sigma_3, \tau_3} 
\int \frac{d{\bf k}_1}{(2\pi)^3} \int \frac{d{\bf k}_2}{(2\pi)^3} 
\int \frac{d{\bf k}_3}{(2\pi)^3} \: n_{k_1} \, n_{k_2} \, n_{k_3}
\nonumber \\[2mm]
&\times f_{\text{R}}^2(p,q) \, 
\langle 1 2 3 \, | \, \vtn \, {\mathcal A}_{123} \, 
| \, 1 2 3 \rangle \,,
\end{align}
where the $n_{k_i}$ denote zero temperature Fermi-Dirac distributions,
$P_{12}$ is the exchange operator for spin, isospin and momenta
of nucleons $1$ and $2$, and $V$ is the volume. 
Moreover, the momentum-conserving delta functions are not
included in the NN and 3N matrix elements.
For the 3N contribution,
\be
f_{\text{R}}(p,q) = \exp \biggl[ - \biggl( \frac{p^2+3 q^2/4}{\la^2}
\biggr)^4 \biggr]
\ee
is the regulator used in the 3N force fits in Ref.~\cite{Vlowk3NF} and
$p$ and $q$ are Jacobi momenta. Exchange terms are included by
means of the antisymmetrizer
\begin{align}
{\mathcal A}_{123} &= (1 + P_{12} P_{23} + P_{13} P_{23}) (1 - P_{23}) \\
&= 1 - P_{12} - P_{13} - P_{23} + P_{12} P_{23} + P_{13} P_{23} \,,
\label{antisym}
\end{align}
where the direct, single-exchange and double-exchange contributions
are apparent in Eq.~(\ref{antisym}). The regulator $f_{\text{R}}(p,q)$
is totally symmetric when expressed in the nucleon momenta
${\bf k}_i$, and thus all exchange terms contain the same regulator
as the direct contribution.

For symmetric nuclear matter, we obtain the following Hartree-Fock
contributions from the leading-order chiral 3N force. For the contact
part ($V_E$) and the $1 \pi$-exchange 3N interaction ($V_D$) we have
\begin{align}
\frac{E^{(1)}_E}{A} &= - \frac{9}{(2 \pi)^4} \, \frac{c_E}{f_\pi^4 \, 
\la_\chi \, \kf^3} \:
\int\limits_0^{2\kf} P^2 dP \int\limits_0^{\sqrt{\kf^2-\frac{P^2}{4}}} 
p^2 dp \int\limits_0^{\kf+\frac{P}{2}} q'^2 dq' \nonumber \\[2mm]
&\times F(p,P;\kf) \, G(q',P;\kf) \, f_{\text{R}}^2(p,2 q'/3)
  \,,
\label{nme} \\[4mm]
\frac{E^{(1)}_D}{A} &= \frac{9}{(2 \pi)^4} \, \frac{g_A \, c_D}{4 
f_\pi^4 \, \la_\chi \, \kf^3} \:
\int\limits_0^{2\kf} P^2 dP \int\limits_0^{\sqrt{\kf^2-\frac{P^2}{4}}} 
p^2 dp \int\limits_0^{\kf+\frac{P}{2}} q'^2 dq' \: \frac{4 p^2}{4 p^2
+ m_\pi^2} \nonumber \\[2mm]
&\times F(p,P;\kf) \, G(q',P;\kf) \, f_{\text{R}}^2(p,2 q'/3) \,,
\label{nmd}
\end{align}
where $F$ and $G$ are phase-space functions
\begin{align}
F(p,P;\kf) &=
\begin{cases}
2 & \text{for $p<\kf-P/2$} \\
2 \: \frac{\textstyle\kf^2-p^2-P^2/4}{\textstyle pP} & \text{otherwise}
\end{cases} \\[2mm]
G(q',P;\kf) &=
\begin{cases}
2 & \text{for $q'<\kf-P/2$} \\
\frac{\textstyle\kf^2-(q'-P/2)^2}{\textstyle q'P} & \text{otherwise} .
\end{cases}
\end{align}
It is useful to have simple estimates for these contributions. If the 
regulator is approximated by unity over the integration region, one 
has $E^{(1)}_E/A = - c_E \, \kf^6 /(12 \pi^4 \, f_\pi^4 \, 
\la_\chi)$. This is reasonable because of the
high power in the exponential regulator $f_{\text{R}}$. We further 
note that in the chiral limit, $m_\pi \to 0$, the ratio of $D$- to $E$-term
contributions is $E^{(1)}_D/E^{(1)}_E = -g_A c_D/4c_E$. 
This is an overestimate for
finite pion mass by $\sim m_\pi^2/\kf^2 \approx 25 \%$.

The remaining $2 \pi$-exchange 3N interaction leads to
\begin{align}
\frac{E^{(1)}_c}{A} &= \frac{9}{(2 \pi)^5} \, \frac{g_A^2}{\kf^3} \:
\int\limits_0^{2\kf} P^2 dP \int\limits_0^{\sqrt{\kf^2-\frac{P^2}{4}}} 
p^2 dp \int\limits_0^{\kf+\frac{P}{2}} q'^2 dq' 
\int\limits_{-1}^1 d\cos \theta_{\bf p} \int\limits_{-1}^1 
  d\cos \theta_{{\bf q}'} 
\int\limits_0^{2\pi} d\varphi \nonumber \\[2mm]
&\times {\mathcal F}(p,P;\kf) \, {\mathcal G}(q',P;\kf) \, 
f_{\text{R}}^2(p,2 q'/3) \nonumber \\[2mm]
&\times \biggl[ - \frac{c_1 m_\pi^2}{f_\pi^4} \biggl(
\frac{{\bf k}_{12} \cdot {\bf k}_{23}}{(k_{12}^2 + m_\pi^2)
(k_{23}^2 + m_\pi^2)} + 2 \, \frac{k_{12}^2}{(k_{12}^2 + m_\pi^2)^2} \biggl)
\nonumber \\[2mm]
&+ \frac{c_3}{2 f_\pi^4} \biggl(
\frac{({\bf k}_{12} \cdot {\bf k}_{23})^2}{(k_{12}^2 + m_\pi^2)
(k_{23}^2 + m_\pi^2)} - 2 \, \frac{k_{12}^4}{(k_{12}^2 + m_\pi^2)^2} \biggl)
\nonumber \\[2mm]
&- \frac{c_4}{2 f_\pi^4} \biggl(
\frac{({\bf k}_{12}\times{\bf k}_{23})^2}{(k_{12}^2+m_{\pi}^2)
(k_{23}^2+m_{\pi}^2)} \biggr) \biggr] \,,
\label{nmc}
\end{align}
where ${\bf k}_{ij} = {\bf k}_i - {\bf k}_j$ and the terms with ${\bf k}_{12}$
only are single-exchanges. Due to the angular dependence of the
$2 \pi$-exchange 3N interaction, the angular integrations do not simplify
and the phase-space restrictions are 
\begin{align}
{\mathcal F}(p,P;\kf) &=
\begin{cases}
1 & \text{for $p<\kf-P/2$} \\
|\cos \theta_{\bf p}| \leqslant 
  \frac{\textstyle \kf^2-p^2-P^2/4}{\textstyle pP} & \text{otherwise}
\end{cases} \\[2mm]
{\mathcal G}(q',P;\kf) &=
\begin{cases}
1 & \text{for $q'<\kf-P/2$} \\
\cos \theta_{{\bf q}'} \leqslant 
  \frac{\textstyle \kf^2-q'^2-P^2/4}{\textstyle q'P} & \text{otherwise} .
\end{cases}
\end{align}
When the regulator is neglected, the $1\pi$- and $2\pi$-exchange 3N force
contributions can be evaluated analytically or can be reduced to
one-dimensional integrals. We give the corresponding expressions in the
Appendix.

\begin{table}[t]
\begin{center}
\begin{tabular}{c|cc}
& $c_D$ & $c_E$ \\ \hline
1.6 & 2.080 & 0.230 \\
1.9 & -1.225 & -0.405 \\
2.1 & -2.062 & -0.625 \\
2.3 & -2.785 & -0.822
\end{tabular}
\caption{Values for the low-energy couplings $c_D$ and $c_E$ adjusted to
the $^3$H and $^4$He binding energies in~\cite{Vlowk3NF}. As discussed in 
the text, the couplings for $\la = 2.1 \fmi$ and $\la = 2.3 \fmi$ have 
been interpolated.}
\label{tab:cdce}
\end{center}
\end{table}

We emphasize that the parameters for the 3N force were adjusted to
the $^3$H and $^4$He binding energies in~\cite{Vlowk3NF}, and
therefore the results presented in this section are
predictions. To obtain results with $\la = 2.1 \fmi$ and $\la = 2.3 \fmi$,
we have interpolated between the smoothly-varying $c_D$ and $c_E$ fit
values.\footnote{For $\la = 2.5 \fmi$, we use the (b) fit of 
Ref.~\cite{Vlowk3NF},
since this lies smoothly between the $\la = 1.9 \fmi$ and $\la = 3.0 \fmi$
fits.} The values for the low-energy couplings $c_D$ and $c_E$ are
listed in Table~\ref{tab:cdce}.
Finally, we note that all Hartree contributions vanish for the 
leading-order chiral 3N interaction due to their spin-isospin structure.

\begin{figure}[t]
\begin{center}
\includegraphics[scale=0.385,clip=]{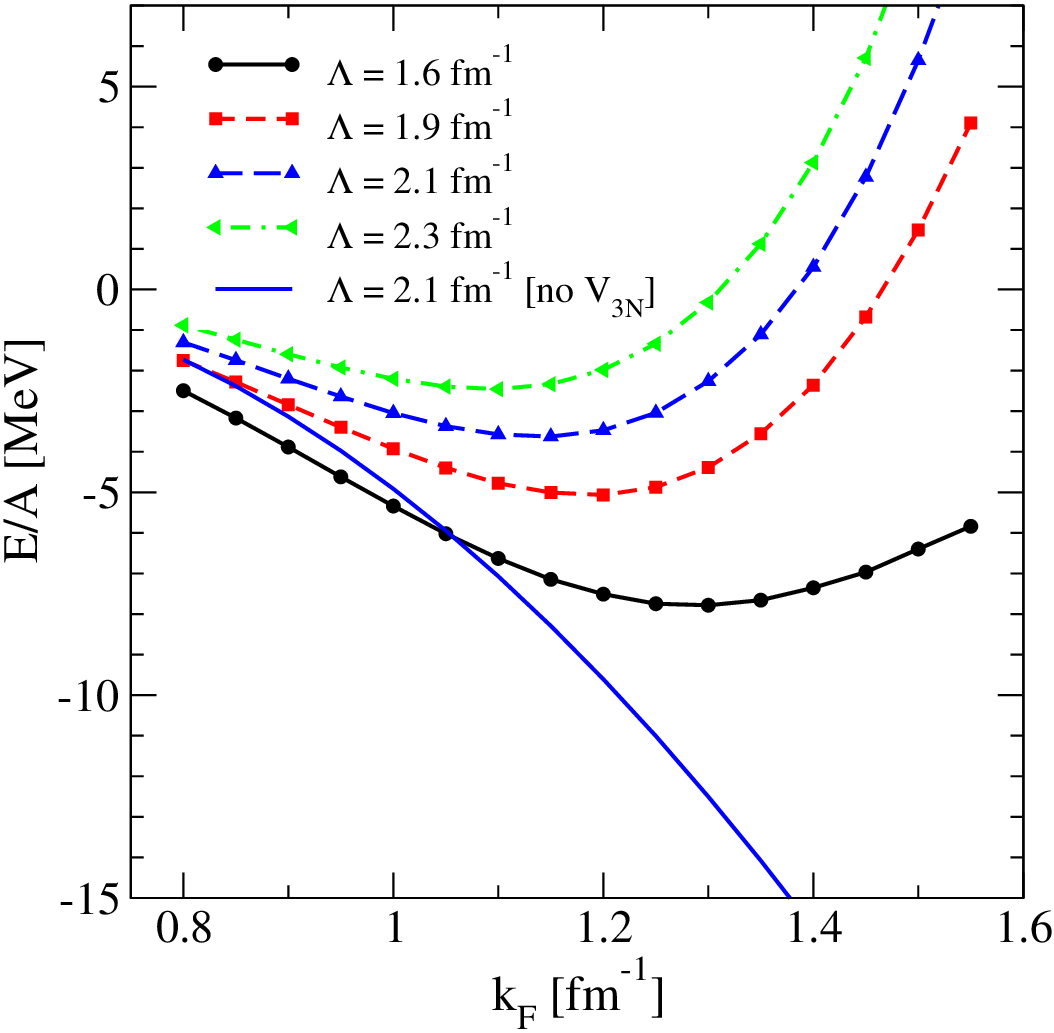}
\includegraphics[scale=0.385,clip=]{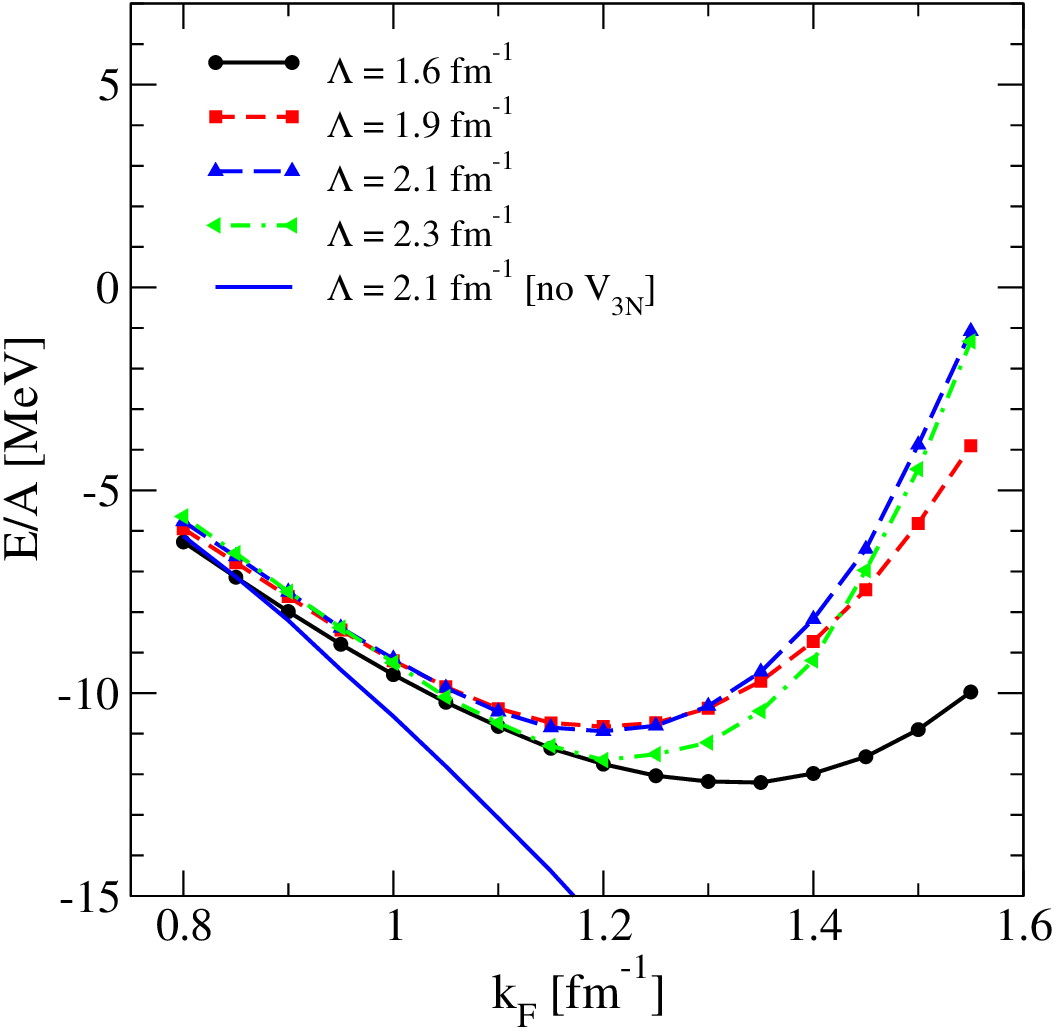}
\end{center}
\caption{Hartree-Fock (left figure) and Hartree-Fock plus dominant 
second-order contributions (right figure) calculated from $\Vlk$
and $\vtn$ for various cutoffs. Details of the approximate second-order 
calculations are given in the text.}
\label{nuclmat}
\end{figure}

In Fig.~\ref{nuclmat}, we present Hartree-Fock results for the energy per 
particle in symmetric nuclear matter calculated from $\Vlk$ and 
$\vtn$ for various
cutoffs. The remaining integrals in Eqs.~(\ref{nme}), (\ref{nmd}), and
(\ref{nmc}) were calculated numerically. The Hartree-Fock results show that
nuclear saturation is due to 3N forces when model-independent, low-momentum NN
interactions are used. As can be seen from the $\Vlk$ results in
Fig.~\ref{nuclmat}, the effects of 3N forces is very small at low 
densities $\kf \lesssim 0.8 \fmi$.
At the Hartree-Fock level, we find a minimum of
$E/A \approx -(2.5 - 8) \mev$ for Fermi momenta $\kf \approx (1.1 - 1.3)
\fmi$ over the cutoff range considered. It is quite promising that nuclear
matter saturates in the Hartree-Fock approximation for low-momentum NN 
and 3N interactions, as conventional NN and 3N interactions require 
complicated nonperturbative treatments to achieve binding.
However, we have not yet formulated a power counting appropriate for finite
density that prescribes these ingredients at leading order. 

As can be seen from Table~\ref{tab:expval},
the dominant 3N contributions are due to the $2 \pi$-exchange
interaction. For $\kf = 1.2 \fmi$ and $\la =1.9 \fmi$, the 3N Hartree-Fock
expectation values are $E^{(1)}_E/A = 1.2 \mev$, 
$E^{(1)}_D/A = -0.8 \mev$ and
$E^{(1)}_c/A = 5.6 \mev$. The repulsive single-exchange $c$-terms 
are more than a factor two larger than the attractive double-exchanges. 
The dominance of the single-exchange terms holds for other densities 
and cutoffs as well.

\begin{table}[t]
\resizebox{5.5in}{!}{
\begin{tabular}{ll|rrrrr|rrrrr}
& \multicolumn{1}{c|}{} & \multicolumn{5}{c|}{Hartree-Fock} & 
\multicolumn{5}{c}{Hartree-Fock + dominant second order} \\
\multicolumn{1}{c}{$\kf$} & \multicolumn{1}{c|}{$\la$} &
\multicolumn{1}{c}{$T$} & \multicolumn{1}{c}{$\Vlk$} &
\multicolumn{1}{c}{$V_{c}$} & \multicolumn{1}{c}{$V_{D}$} &
\multicolumn{1}{c|}{$V_{E}$} & \multicolumn{1}{c}{$T$} &
\multicolumn{1}{c}{$\Vlk$} & \multicolumn{1}{c}{$V_{c}$} & 
\multicolumn{1}{c}{$V_{D}$} & \multicolumn{1}{c}{$V_{E}$} \\ \hline
$1.0$ & $1.6$ & $12.44$ & $-19.62$ & $1.65$ & $0.42$ & $-0.22$ &
$15.50$ & $-26.58$ & $1.49$ & $0.34$ & $-0.29$ \\
 & $1.9$ & $12.44$ & $-18.18$ & $1.67$ & $-0.25$ & $0.40$ &
$16.29$ & $-26.81$ & $0.85$ & $-0.09$ & $0.55$ \\
 & $2.1$ & $12.44$ & $-17.35$ & $1.67$ & $-0.42$ & $0.62$ &
$16.92$ & $-27.04$ & $0.11$ & $0.05$ & $0.79$ \\
 & $2.3$ & $12.44$ & $-16.56$ & $1.67$ & $-0.56$ & $0.81$ &
$17.60$ & $-27.27$ & $-0.89$ & $0.43$ & $0.85$ \\ \hline 
$1.2$ & $1.6$ & $17.92$ & $-31.47$ & $5.37$ & $1.31$ & $-0.64$ &
$20.86$ & $-37.66$ & $4.59$ & $1.03$ & $-0.65$ \\
 & $1.9$ & $17.92$ & $-28.95$ & $5.61$ & $-0.81$ & $1.18$ &
$21.80$ & $-37.38$ & $3.99$ & $-0.50$ & $1.28$ \\
 & $2.1$ & $17.92$ & $-27.51$ & $5.67$ & $-1.37$ & $1.84$ &
$22.87$ & $-37.53$ & $2.27$ & $-0.37$ & $1.82$ \\
 & $2.3$ & $17.92$ & $-26.13$ & $5.70$ & $-1.86$ & $2.42$ &
$24.32$ & $-37.95$ & $-0.38$ & $0.51$ & $1.78$ \\ \hline
$1.35$ & $1.6$ & $22.67$ & $-42.47$ & $10.75$ & $2.59$ & $-1.21$ &
$26.09$ & $-47.85$ & $8.73$ & $1.96$ & $-1.12$ \\
 & $1.9$ & $22.67$ & $-38.82$ & $11.95$ & $-1.69$ & $2.34$ &
$26.75$ & $-46.72$ & $9.14$ & $-1.16$ & $2.24$ \\
 & $2.1$ & $22.67$ & $-36.74$ & $12.19$ & $-2.91$ & $3.68$ &
$28.05$ & $-46.47$ & $6.99$ & $-1.33$ & $3.22$ \\
 & $2.3$ & $22.67$ & $-34.77$ & $12.30$ & $-3.97$ & $4.89$ &
$30.06$ & $-46.45$ & $3.10$ & $-0.35$ & $3.26$
\end{tabular}}
\vspace*{2mm}
\caption{Expectation values of the kinetic energy ($T$), 
$\Vlk$ and the different $\vtn$ contributions in MeV. The expectation
values are obtained with the Feynman-Hellman method, and $\la$ and $\kf$
are given in $\fmi$.}
\label{tab:expval}
\end{table}

Next, we compute the dominant second-order contributions to the 
energy per particle.
The approximate second-order calculation is carried out in two
steps. First, we convert the 3N force into a density-dependent NN
interaction $\veff$ by summing the third particle over occupied states in
the Fermi sea,
\be
\langle 1 2 \, | \, \veff  \, | \, 1' 2' \rangle = 
\tr_{\sigma_3,\tau_3} \int \frac{d{\bf k}_3}{(2\pi)^3}
\: n_{k_3} f_{\text{R}}(1 2 3) \, f_{\text{R}}(1' 2' 3) \,
\langle 1 2 3 \, | \, \vtn \, | \, 1' 2' 3 \rangle .
\ee
A first-order calculation of $E^{(1)}$ using $\veff$ includes the 
single-exchange contributions of the Hartree-Fock 
calculation with $\vtn$.
We then calculate the second-order contributions $E^{(2)}$ from 
$\Vlk$ plus density-dependent $\veff$,
\begin{align}
\frac{E^{(2)}}{V} &= - \frac{1}{4} \: \prod_{i=1}^4 \, \biggl(
\tr_{\sigma_i,\tau_i} \int\frac{d{\bf k}_i}{(2\pi)^3} \biggr) \: 
n_{k_1} \, n_{k_2} \, (1-n_{k_3}) \, (1-n_{k_4}) \nonumber \\
& \times \frac{| \langle 1 2 \, | \, (\Vlk+\veff) \, (1-P_{12}) \,
| \, 3 4 \rangle|^2}{\epsilon_{k_3}+\epsilon_{k_4}
-\epsilon_{k_1}-\epsilon_{k_2}} \, (2 \pi)^3 \, \delta^{(3)}({\bf k}_1
+ {\bf k}_2-{\bf k}_3-{\bf k}_4) \,.
\end{align}
For the intermediate-state integrations, the phase-space is angle-averaged 
and we use a continuous spectrum for $\epsilon_{k} = k^2/(2 m^*)$. The
angle-averaging approximation is expected to be reliable~\cite{Suzuki}.
Here the effective mass is determined from the first-order
self-energy correction from $\Vlk + \veff$ at the Fermi surface,
\be
\frac{m^*}{m} = \biggl( 1-\frac{k}{m} \, \frac{\partial \Sigma(k)}{
\partial k} \biggr|_{k=\kf} \, \biggr)^{-1} \,,
\ee
where the spin-isospin independent part of the self-energy $\Sigma(k)$
is given by
\be
\Sigma(k) = \frac{1}{4} \: \tr_{\sigma_1,\tau_1} \, \tr_{\sigma_2,\tau_2}
\int\frac{d{\bf k}_2}{(2\pi)^3} \: n_{k_2} \, 
\langle 1 2 \, | \, (\Vlk+\veff) \, (1-P_{12}) \, | \, 1 2 \rangle \,.
\ee
In this approximation, we find for the effective mass $m^*/m = 0.72$, 
$0.67$ and $0.65$ for $\la = 1.9 \fmi$ at $\kf = 1.0 \fmi$, $1.2 \fmi$ 
and $1.35 \fmi$ respectively.

\begin{figure}[t]
\begin{center}
\includegraphics[scale=0.45,clip=]{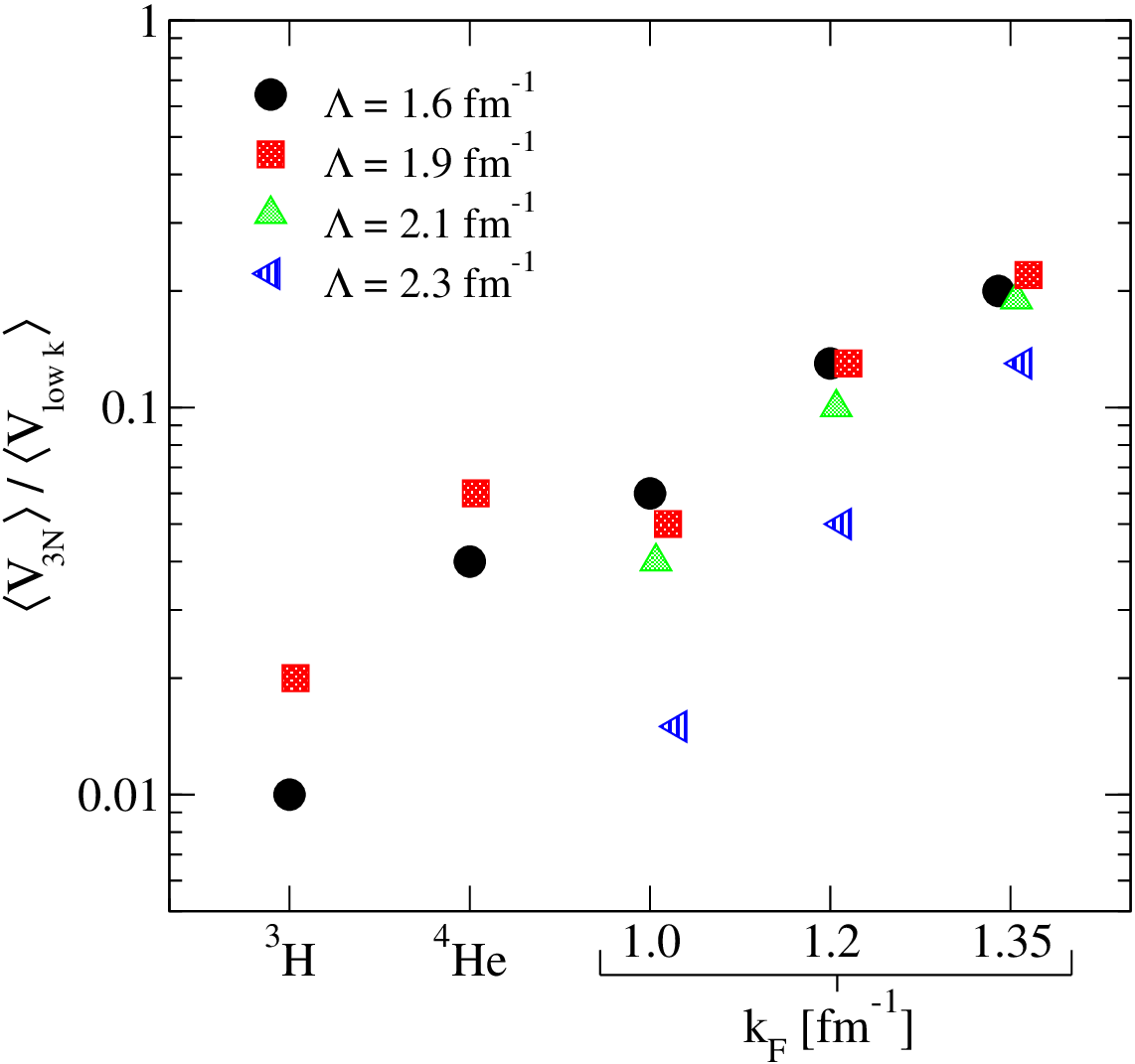}
\end{center}
\caption{Ratio of $\vtn$ to $\Vlk$ expectation values in the approximate
second-order calculation. The expectation values are obtained by the 
Feynman-Hellman method. For nuclear matter we take for $\langle \vtn
\rangle$ the total 3N contribution, whereas the values shown for
$^3$H and $^4$He are the largest 3N contributions (because of
cancellations in the light nuclei).}
\label{expval}
\end{figure}

Our nuclear matter results including these dominant second-order 
contributions are shown in the right panel of Fig.~\ref{nuclmat} 
and are promising in a number of aspects. First, we observe that 
the second-order corrections move the saturation curves towards 
the empirical value. With the additional attraction, we find a 
minimum of $E/A \approx - (11 - 12) \mev$ for Fermi momenta $\kf 
\approx (1.2 - 1.35) \fmi$ over the cutoff range considered. 
Third-order particle-particle and hole-hole contributions are
found to be small ($\lesssim 1 \mev$). Second, 
the cutoff dependence is dramatically reduced when second-order 
contributions are included. We note that the determination 
of the $c_i$ are not unique~\cite{N3LO,const,Meissner1,Meissner2}. 
For example, the values $c_3 = -3.2 \gevi$ and $c_4 = 5.4 \gevi$
determined by Entem and Machleidt~\cite{N3LO} 
would yield an additional $\sim 1\mbox{--}2 \mev$ binding. 
We also emphasize that our calculation
neglects second-order double-exchange contributions, which are
expected to be attractive and nearly cutoff-independent. 
Furthermore, the effective mass
approximation needs to be improved to treat the full momentum dependence
of the Hartree-Fock self-energy.

Our results establish that the 3N force drives
saturation for low-momentum interactions. However, this 
does not imply that the 3N contributions are unnaturally large. 
To provide further insight, we list in Table~\ref{tab:expval} 
the expectation values for the kinetic and potential energies 
in symmetric nuclear matter. The expectation values are obtained
by the Feynman-Hellman method, where one multiplies the operator
of interest in the Hamiltonian by a parameter $g$, so that $H^g =
(H-O) + g O$, and computes $\langle O \rangle = (d \langle
H^g \rangle / dg) |_{g=1}$. The cutoff dependence of the expectation
value for the kinetic energy $\langle T \rangle$ reflects the fact
that correlations in the wave functions are not observable and are reduced
as the cutoff is lowered. Note that $\langle T \rangle$ is significantly
larger for large-cutoff interactions, see e.g.,~\cite{AKMAL}.

We also show in 
Fig.~\ref{expval} the ratio of $\langle \vtn \rangle / \langle
\Vlk \rangle$ for the triton, alpha particle, and nuclear matter 
at various densities. According to chiral power-counting estimates, 
a natural 
3N contribution should scale as $\langle \vtn \rangle \sim (Q/\la)^3 \,
\langle \VNN \rangle$, where $Q$ is a typical momentum
scale of the low-energy system. Our results for the expectation 
values show that the 3N contributions are not unnatural; for
example, the ratio $\langle V_{3N} \rangle / 
\langle \Vlk \rangle \, (\la/\kf)^3$ is $0.31$, $0.51$, $0.53$, 
and $0.35$ for the second-order calculation at $\kf = 1.2 \fmi$
and $\la = 1.6 \fmi$, $1.9 \fmi$, $2.1 \fmi$, and $2.3 \fmi$ 
respectively.
Nevertheless, the role of 3N forces in driving saturation 
means that we will ultimately
want to identify a power counting for finite density that includes
both NN and 3N Hartree-Fock contributions at leading order. 
It is also interesting to note that the deviation from the empirical 
saturation point is compatible with a 3N or 4N force 
contribution of order $(Q/\lm)^4$.

Finally, as was previously 
found for the binding energies of $A=3,4$ nuclei~\cite{Vlowk3NF}, 
we find that the 3N force may also be perturbative in 
nuclear matter calculations for cutoffs $\la \lesssim 2 \fmi$. 
For example, the expectation value of $\langle \Vlk \rangle$ 
for $\kf = 1.2 \fmi$ in the absence of 3N forces are
$\langle \Vlk \rangle = -37.5 \mev$, $-37.9 \mev$, $-38.2
\mev$, and $-38.5 \mev$ for $\la = 1.6 \fmi$, $1.9 \fmi$,
$2.1 \fmi$, and $2.3 \fmi$ respectively. Comparing
these with the expectation values in Table~\ref{tab:expval}, we
find that 3N effects in the wave functions are small. In contrast, 
the expectation value of the Argonne $v_{18}$ potential changes 
by $1.6 \mev$ after resummations when the 
Urbana IX 3N interaction is included~\cite{AKMAL}.

\section{Conclusions and Summary}

In this paper, 
the nonperturbative nature of inter-nucleon interactions is explored
by varying the momentum cutoff of a two-nucleon potential.
Conventional force models, which have large cutoffs, 
are nonperturbative because
of strong short-range repulsion, the iterated tensor interaction, and
the presence of bound or nearly-bound states.
But for low-momentum interactions with
cutoffs around $2 \fmi$, the softened potential
combined with Pauli blocking leads to 
corrections in nuclear matter
in the particle-particle channel that are well converged at
second order in the potential, 
suggesting that perturbation theory (as in a loop expansion) 
can be used in place of Brueckner resummations. 
A Weinberg eigenvalue analysis provides quantitative backing to these
observations.

We present the first
calculations of nuclear matter using the low-momentum
two-nucleon $\Vlk$ with a corresponding 3N force
from chiral EFT, 
which are fit in free space and to binding energies of three-nucleon
systems with no adjustments when applied at finite density.
This combination
exhibits nuclear binding  
in the Hartree-Fock approximation and becomes 
significantly less cutoff dependent
with the inclusion of the dominant second-order contributions. 
The role of the 3N force
is essential to obtain saturation, but the contribution to the total
potential energy is still compatible with EFT power-counting estimates.

At lower cutoffs around $2 \fmi$,
the iterated tensor interaction in
the two-body sector does not play a major role in nuclear saturation,
in contrast to the conventional wisdom.
We emphasize, however, that the relative importance of contributions to
observables from the tensor force or from three-body forces are scale
or resolution dependent.
Renormalizing the potential to a different cutoff redistributes these
contributions.  Our point is that potentials
with low-momentum cutoffs may be
superior for practical nuclear many-body calculations.

We further 
emphasize that the use of $\Vlk$ is \emph{not} a replacement for EFT
but can be used to improve EFT interactions for many-body applications.
The strategy is to start with a consistently truncated
chiral EFT (including many-nucleon interactions), using the highest feasible
cutoff $\Lambda$ to minimize the truncation error in matching.
Then we evolve the EFT to lower cutoffs (lower resolution) using 
the RG.  This running does not, in the present case, 
involve large separations of scale 
that would justify a truncated derivative expansion, 
so we keep induced operators to all orders to
prevent the truncation error from growing.  

One could be misled into thinking a higher cutoff implies that the
physics is more valid.  As stressed by Lepage,  one should match 
a low-energy theory
to data or an underlying theory with a cutoff close to the scale at
which unknown physics starts to be resolved \cite{LEPAGE}. 
There is no advantage to
putting it higher, as it can be counterproductive because
counterterms (which translates into the potential here) must cancel
loop contributions from intermediate states that are incorrectly
represented.\footnote{In other words, the model-dependent cores in
the conventional NN potentials mandate model-dependent parts in the
3N force to counteract the effects of the former.} 
But this does not mean that EFT calculations cut off at the
breakdown scale are optimal for applications to nuclei. Instead, one
should choose a resolution appropriate to the problem by 
matching the chiral EFT at the breakdown scale ($\Lambda \approx
3\mbox{--}4 \fmi$) and then evolve downward. We find that
the decrease in resolution to $\Lambda \sim 2 \fmi$ is particularly
important for reducing the tensor force.

The application of chiral EFT in perturbation theory by Lutz
{\it et al.}~\cite{Lutz} and Kaiser
{\it et al.}~\cite{Kaiser} may be justified in part by our results,
although further detailed comparisons are needed.
We expect, however, that the convergence will be improved for smaller
cutoffs ($\la = 3.28 \fmi$ in~\cite{Kaiser}). 
In contrast to our results, which are
predictions based on potential fits to scattering observables
with relatively small residual cutoff dependence of observables,
Kaiser {\it et al.} fine tune the cutoff to simulate the effects of 
omitted contact terms. Contact interactions are included
in $\Vlk$ and their contributions to nuclear matter are important.
In addition, $2 \pi$-exchange 3N
contributions, which are essential for saturation in
the present approach, are not included in Ref.~\cite{Kaiser}.
It is conceivable that a substantial part of our 3N force
contributions are captured by the explicit Delta isobars 
included in Fritsch {\it et al.}~\cite{Fritsch}, since the low-energy 
constants $c_3$ and $c_4$ are to a large extent saturated by the Delta 
isobar~\cite{BKM}. Therefore, a detailed comparision with this 
work will be very interesting for the future.

The present calculations can be improved in several respects, 
which is reflected in 
residual cutoff dependence of the energy per particle in nuclear matter. 
First,
the current calculations are not complete at second order in many-body
perturbation theory.  We expect
that the missing second-order 
contributions are, in fact, more important than omitted
higher-order contributions.  
The double-exchange $c$-terms are expected to give attractive and 
nearly cutoff-independent contributions. Thus,
a full second-order calculation is a high priority. 
Other improvements would be a better treatment of self-consistency 
at second order.
In parallel, we will use the cutoff dependence of $\Vlk$ as a
tool to develop a power counting at finite density that specifies
the appropriate truncations and justifies our conclusions.

With the present three-body force, 
a better determination of the $c_i$ coefficients is needed,
since $c_3$ and $c_4$ contribute at the MeV level.
Of course, it will be important to eventually evolve the many-body
forces to lower cutoffs at the same time as the two-nucleon force is
renormalized, as this will generate better estimates of higher-order 
three-body and four-body contributions.
In addition, the impact of using cutoffs for $\Lambda < 2\kf$ needs
further investigation.

Finally,
there are particle-hole contributions to consider. These contribute
to the energy at third order. 
While we have shown that nuclear matter
is perturbative in the particle-particle channel and that 3N
forces naturally provides a saturation mechanism, there is still the
possibility that a nonperturbative summation in the particle-hole 
channels is required~\cite{Jackson}.
The lore with
conventional potentials is that these correlations are important for the
nuclear response but not for the energy of bulk nuclear matter.
If this also holds for low-momentum potentials, 
then we can conclude that nuclear matter under ordinary conditions can
be treated perturbatively in the inter-nucleon forces.

For applications to astrophysics, however,
we may seek higher densities (e.g., for applications to the neutron star
equation of state).
The increase in three-body contributions with density above
saturation implies that one should
only do controlled extrapolations to higher density.
Controlled extrapolations are also needed to reliably describe
neutron- or proton-rich nuclei towards the drip lines.
The use of a variable cutoff and natural estimates of omitted
contributions is the ideal tool for such investigations.

\begin{ack}
We are grateful to Norbert Kaiser for pointing out the $c_4$ Fock
contribution.
We thank Gerry Brown, Bengt Friman, Chuck Horowitz, Brian Serot 
and Bira van Kolck for useful comments and discussions.
This work was supported in part by the National Science Foundation
under Grants No.~PHY--0098645, No.~PHY--0244822, and No.~PHY--0354916,
and by the US Department of Energy under Grants No.~DE-FG02-87ER40365,
DE-FC02-01ER41187 and DE-FG02-00ER41132.
\end{ack}

\appendix
\section{Pionic three-nucleon force contributions without regulator}

In this Appendix, we provide analytical expressions for the $1 \pi$-
and $2 \pi$-exchange 3N force contributions at the Hartree-Fock level,
when the regulator $f_{\text{R}}$ is neglected. Analogous $2 \pi$-exchange 
contributions with explicit Delta isobars have been evaluated 
in~\cite{Fritsch}. This provides an estimate for these 3N contributions
and a check for the numerical integration with regulator.

For the $1 \pi$-exchange 3N interaction ($V_D$), we obtain from 
Eq.~(\ref{nmd}) with $f_{\text{R}}=1$
\begin{align}
\frac{E^{(1)}_D}{A} &= \frac{g_A \, c_D}{f_\pi^4 \, \la_\chi}
\, \frac{3}{288 \, (2 \pi)^4} \, \biggl[ \, 32 \, \kf^6 - 72 \, \kf^4 
\, m_\pi^2 + 12 \, \kf^2 \, m_\pi^4 \nonumber \\[2mm]
&+ 96 \, \kf^3 \, m_\pi^3 \, \arctan\biggl(\frac{2 \kf}{m_\pi}
\biggr)-\bigl(36 \, \kf^2 \, m_\pi^4 + 3 \, m_\pi^6\bigr) \log\biggl(1+
\frac{4 \kf^2}{m_\pi^2}\biggr)\biggr] \,,
\end{align}
and for the $2 \pi$-exchange 3N interaction ($V_c$), 
Eq.~(\ref{nmc}) leads to
\begin{align}
\frac{E^{(1)}_c}{A} &= \frac{g_A^2 \, m_\pi^2 \, c_1}{\kf^3 \, f_\pi^4}
\, \frac{9}{2 (2 \pi)^4} \, \biggl[ \, \frac{\kf^3}{3 m_\pi} \,
\partial_{m_\pi} \biggl( m_\pi^6 \int\limits_0^u x dx \, G_s(x,u) \biggr) 
\nonumber \\[2mm]
&+ \frac{m_\pi^7}{128} \: \int\limits_0^u x^2 dx \, G_v^2(x,u) \biggr]
\nonumber \\[2mm]
&+ \frac{g_A^2 \, c_3}{\kf^3 \, f_\pi^4}
\, \frac{9}{4 (2 \pi)^4} \, \biggl[ \, \frac{m_\pi^9}{72} \:
\int\limits_0^u dx \, \bigl( 3 \, G_s^2(x,u) + G_t^2(x,u) \bigr)
\nonumber \\[2mm]
&- \frac{2 \, \kf^3 \, m_\pi^6}{3} \: \int\limits_0^u x dx \, G_s(x,u)
- \frac{\kf^3 \, m_\pi}{3} \, \partial_{m_\pi} \biggl( m_\pi^6 
\int\limits_0^u x dx \, G_s(x,u) \biggr) 
\nonumber \\[2mm]
&+ \frac{g_A^2 \, m_\pi^9 \, c_4}{\kf^3 \, f_\pi^4}
\, \frac{3}{(4 \pi)^4} \: \int\limits_0^u dx \, \bigl( G_t^2(x,u) - 
G_s^2(x,u) \bigr) \biggr] \,.
\end{align}
Here $u=\kf/m_\pi$ and the auxiliary functions $G_s(x,u)$, $G_v(x,u)$
and $G_t(x,u)$ are defined as
\begin{align}
G_s(x,u) &= \frac{4ux}{3} \, (2u^2-3) + 4x \bigl(\arctan(u+x) + \arctan(u-x)
\bigr) \nonumber \\[2mm]
&+ (x^2-u^2-1) \log\frac{1+(u+x)^2}{1+(u-x)^2} \nonumber \\[2mm]
G_v(x,u) &= \frac{1}{x^2} \, \biggl( \bigr( u^4-2u^2(x^2-1) +(x^2+1)^2 \bigr)
\log\frac{1+(u+x)^2}{1+(u-x)^2} \nonumber \\[2mm]
&- 4ux \, (x^2+u^2+1) \biggr) \nonumber \\[2mm]
G_t(x,u) &= \frac{ux}{6} \, (8u^2+3x^2) - \frac{u}{2x} \, (1+u^2)^2
\nonumber \\[2mm]
&+ \frac{1}{8} \, \biggl( \frac{(1+u^2)^3}{x^2} - x^4 + (1-3u^2)(1+u^2-x^2) 
\biggr) \log\frac{1+(u+x)^2}{1+(u-x)^2} \,.
\end{align}


\end{document}